\documentstyle[psfig,11pt,aaspp4]{article}

\newcommand{\etal}{{\it{et al.}}~}
\newcommand{\ie}{{\it{i.e.}}~}
\newcommand{\eg}{{\it{e.g.}}~}

\lefthead{Rizza et al.}
\righthead{X-ray and Radio Interactions in Cooling Flows }

\begin{document}

\title{X-ray and Radio Interactions in the Cores of Cooling Flow Clusters }

\author{E. Rizza,\altaffilmark{1} C. Loken, M. Bliton,\altaffilmark{1} K. Roettiger, J.O. Burns}
\affil{Department of Physics and Astronomy, University of Missouri, Columbia, MO 65211}

\and

\author{F. N. Owen}
\affil{National Radio Astronomy Observatory\altaffilmark{2}, Socorro, NM 87801}

\altaffiltext{1}{Department of Astronomy, New Mexico State University, Las Cruces, NM 88003-8001} 
\altaffiltext{2}{The National Radio Astronomy Observatory is operated by 
Associated Universities, Inc., under a cooperative agreement with the National Science Foundation}

\begin{abstract}
We present high resolution ROSAT x-ray and radio observations of three
cooling flow clusters containing steep spectrum radio sources at their
cores. All three systems exhibit strong signs of interaction between
the radio plasma and the hot intracluster medium. Two clusters, A133
and A2626, show enhanced x-ray emission spatially coincident with the
radio source whereas the third cluster, A2052, exhibits a large region
of x-ray excess surrounding much of the radio source.  Using 3-D
numerical simulations, we show that a perturbed jet propagating
through a cooling flow atmosphere can give rise to amorphous radio
morphologies, particularly in the case where the jet was ``turned
off'' and allowed to age passively. In addition, the simulated x-ray
surface brightness produced both excesses and deficits as seen
observationally.

\end{abstract}


\section{Introduction}
Several classes of radio sources are known to exhibit morphological
and physical properties which are dramatically influenced by their
environments, \eg amorphous radio sources (Burns 1990; Burns 1996),
wide angle tailed radio sources (Pinkney 1995; Gomez \etal 1997), and
narrow angled tail radio sources (Bliton \etal 1998). Consequently,
the most profoundly affected sources are generally found in the most
extreme environments, \ie within clusters of galaxies.  Evidence
suggests, however, that the interaction between the radio source and
the intracluster medium (ICM) not only affects the radio source, but
also impacts the surrounding medium in an observationally and at
times, dynamically significant manner.

One such case is in the powerful classical double Cygnus
A. Observations of cavities and enhancements in the cluster x-ray
emission have been interpreted as the result of sweeping and
compression of the material by the expanding lobes and bow shock,
respectively (Carilli, Perley \& Harris 1994). Clark, Harris and
Carilli (1997) lent support to this hypothesis through a series of 3D
hydrodynamical simulations designed to mirror the observable
conditions of CygA.

A2199 is another example where the radio source (3C338) and x-ray
emission seem to have substantial mutual influence on one
another. Owen \& Eilek (1998) suggest that the radio source may have
had a significant effect on the energetics of the cluster core,
creating a complex environment which differs from a symmetric cooling
flow at small radii. They interpret the radio source as having
severely disrupted due to conditions in the cooling core, producing
diffuse, bubble structures as a result. Other examples displaying
similar evidence of a dynamical interaction between a central radio
source and the ambient cooling flow media have been detected in the
inner region of NGC1275 (3C 84 in the Perseus cluster; Bohringer
\etal 1993) and M87 (Bohringer \etal 1995). New, high resolution radio
maps of M87 show that the situation may become even more complex,
including outflow conditions in the core (as opposed to a cooling
inflow), mixing of the radio and x-ray plasma, and possible periodic
cycling of the radio source (Owen, Eilek \& Kassim 1999).

The developing complexity of these systems invites further
investigation of radio sources residing at the cores of clusters of
galaxies. Through observational analysis, paired with numerical
simulations, information about both the radio source and the cluster
environment can be extracted and models for their
interactions/connections refined. Steep spectrum radio sources, often
found at the cores of cooling flow clusters (Burns \etal 1997), are
believed to be under-pressured and confined by the surrounding
medium. Such confinement causes the amorphous shapes in the sources
and a steepening of their spectral index due to aging of the electron
population responsible for the synchrotron emission while maintaining
surface brightness against adiabatic losses. These sources, therefore,
serve as particularly good signposts for an x-ray/radio interaction.

In this paper, we present new high resolution ROSAT x-ray and VLA
radio observations of three cooling flow clusters containing steep
spectrum radio sources at their cores. These systems exhibit strong
signs of interaction between the radio plasma and the hot intracluster
medium. Two clusters, A133 and A2626, show enhanced x-ray emission
spatially coincident with the radio source whereas the third cluster,
A2052, has a large region of x-ray excess which surrounds much of the
steep spectrum radio emission.  We compare the observed features with
results from 3D numerical simulations of perturbed jets propagating
through a cooling flow atmosphere. With these simulations we are able
to reproduce x-ray cavities and enhancements through compression and
sweeping of the ICM.  However, the morphology of the numerical jet
does not exhibit the extreme bending which we see in the observed
clusters implying that some other mechanism, beyond the initial
perturbation, must be involved in order to break the symmetry of the
jet. A preliminary analysis of this work was presented in Burns \etal
(1997).

The paper is organized as follows. In \S 2 we define our sample and in
\S 3 we describe the radio observations and morphologies of the sources. 
The x-ray observations and reductions are presented in \S 4 and the
x-ray substructure analysis in \S 5. We discuss the possible emission
mechanisms for the x-ray excesses in the observational data and
determine magnetic field estimates for the inverse Compton case in \S
6.  Section 7 contains a description of the model atmosphere and jet
used in our numerical simulations and \S 8 contains the discussion and
comparisons between the observed and modeled data. Throughout the
paper, $\rm H_{0}= 75$ $\rm km/s/Mpc$ and $q_{0} = 0.5$ unless stated
otherwise.

\section{The sample}
The clusters for this study comprise a subsample of those selected in
Ge \& Owen (1994). Their sample was defined to include clusters which
(1) are nearby and contained in the Abell cluster catalog; (2) are
known to possess cooling flows with an accretion rate greater than $80
M_{\odot} yr^{-1}$ (\cite{arn88}); and (3) contain an extended radio
source which is near the center of the cooling flow and powerful
enough to be observed with the VLA. These clusters were
cross-referenced with x-ray data available from ROSAT High Resolution
Imager (HRI), leaving three interesting candidates. We have our own
HRI data for A133 and A2626 and the data for A2052 were gathered from
the public archives. Though selected from the Arnaud (1988) catalog,
more recent studies have determined mass accretion rates largely
consistent with the these results. These clusters all have relatively
strong cooling flows with rates of $110^{+71}_{-67}$, $94^{+84}_{-37}$
and $53^{+36}_{-30}$ $M_{\odot} yr^{-1}$ ($\rm H_0 = 50$) for A133,
A2052 and A2626 respectively (White, Jones \& Forman 1997).

\section{Radio observations and source morphology}
The radio data for the sample were observed using the Very Large Array
(VLA) in several different modes including new, 20-cm radio
observations of A133 and A2626. The dates of the observations are
listed in Table 1 along with the array configuration, bandwidth and
total flux density for both sources.  Secondary flux calibrators were
observed every 20 minutes in order to account for phase errors and
changes in the atmospheric and system conditions.  The absolute flux
of the radio sources was calibrated by observing the primary flux
calibrator 3C48 at the beginning and end of the runs.  The data were
processed using standard procedures within the {\tt AIPS} environment
where the maps were created using the task {\tt IMAGR}, after cleaning
and self-calibrating the data. A133 was observed in two arrays in
order to increase the sensitivity to smaller scale structures and the
visibility data were combined via the task {\tt DBCON}. The radio maps
shown in Fig. 1-2 were restored using a Gaussian beam of $9.7''\times
4.9''$ and $12.9''\times 12.1''$ for A133 and A2626, respectively. The
asymmetry in the beam for A133 is caused by the low declination of the
source.

The morphology of both these sources is very complex. They are known
to be in the class of steep spectrum radio sources and thus their
amorphous appearance is expected. Previous observations of A133
revealed it to have an extremely steep spectrum, particularly in the
northern, elongated region ($\alpha = 2.8$, where $S_{\nu} \propto
\nu^{-\alpha}$; \cite{slee84}). The compact component to the south is
associated with the cD galaxy at the center of the cluster at a
redshift of z=0.057. Earlier observations suggested that the two
components (core and diffuse) were detached from one another. This
implied the steep spectrum emission was a fossil radio source that was
either an ejected plasmoid from the central source or the remnant of
an accreted radio galaxy (\cite{slee84}). The observations presented
here show the two components to be clearly connected and thus a
continuous system.

Abell 2626 is a very unusual source exhibiting a compact, unresolved
core, and two steep spectrum, diffuse structures in a z-shaped
orientation with respect to the central source. The compact component
is associated with the centrally dominant elliptical galaxy
(\cite{owen95}) and the diffuse emission has no optical
counterpart. The symmetry of these components and their steep spectral
index ($\alpha = 1.97 2.0$, Komissarov \& Gubanov (1994)) imply they
are spectrally aged lobes from the central source. A dynamical
analysis of Abell 2626 revealed it to be a surprisingly complex
cluster despite the symmetry of the large scale x-ray
distribution. Mohr, Geller \& Wegner (1996) found the galaxy
distribution to be composed of three subgroups with indications of an
ongoing merger between two of the components.

The radio data for Abell 2052 (3C 317), kindly supplied by J.-H. Zhao,
are a combination of A and B configurations at 20-cm (see
\cite{zhao93} for details). This source also has a very peculiar
morphology with a compact core and a twisting, diffuse outer
component. The outer halo was found to have an average spectral index
of $\alpha = 1.5$ at $\nu <0.5$ GHz. Optical emission line gas,
coincident with the weaker regions of the radio emission, has also
been detected for A2052 (Heckman \etal 1989; Zhao \etal 1993).

\section{X-ray reductions and analysis}
All three clusters were observed with the High Resolution Imager
aboard the ROSAT x-ray satellite.  Two of these clusters, A133 and
A2052, also have public ROSAT PSPC data which were used to supplement
the HRI data.  Table 2 lists the cluster name, the sequence number,
and the Abell cluster position for each source. The exposure times and
redshift for each cluster are also listed.

All x-ray reductions were performed using standard {\tt IRAF/PROS}
software. The HRI data were first examined and edited for intervals of
high spurious count rates. With these time intervals excluded, the
data were binned into $3''$ pixel images to allow for an increase in
the signal to noise without significantly degrading the resolution of
the instrument. Background values for each cluster were determined
from circular annuli selected outside the cluster emission and were
subtracted from the original images.  Lastly, the images were divided
by the edited live times in order to generate count rate maps. The
flat response of the HRI provides for less than a few percent error
using this method. It should be noted that although these final images
were utilized for part of the analysis, simple count maps were
employed for the substructure tests (see \S 5). The HRI contour maps
were overlaid onto optical grey scales for the three clusters and are
shown in Figs. 1-3.

The HRI observations are only sensitive to the inner $\approx 100$ kpc
of the cluster emission, essentially the core regions. For the purpose
of this study, however, the emission extends far beyond the regions
surrounding the radio sources in all three cases. The overall x-ray
morphology of the three clusters is relatively relaxed outside the
inner $\sim 30-40$ kpc.  Abell 133 shows a strong central peak in the
x-ray surface brightness consistent with the presence of a cooling
flow, yet this peak is offset to the south relative to the outer
emission. The apparent asymmetry can be attributed to the excess
emission coincident with the steep spectrum (diffuse) region of the
radio source. Abell 2626 also possesses a smooth, rather symmetric
x-ray distribution at radii greater than about 45 kpc, with an
elongation coincident with the steep spectrum, diffuse radio
component. Of the three cases, the core x-ray emission for Abell 2052
is the most complex. The central few arcsec shows a flat, V-shaped
morphology, becoming increasingly elliptical at larger radius.

To supplement the high resolution HRI data with sensitivity to low
surface brightness emission, the PSPC data were retrieved for the two
publicly available clusters, A133 and A2052. The hard band images
(0.1-2.4 keV) were background subtracted using values determined from
regions outside the cluster emission yet still interior to the inner
ring of the instrument. They were then exposure corrected using the
default exposure maps to produce count rate maps. These surface
brightness images from the PSPC confirm the larger-scale, relaxed
morphology implied by the HRI.

\section{X-ray substructure analysis}
The first test for substructure employed in this analysis was
motivated by the desire to accentuate observational signatures of
possible interactions between the x-ray and radio emission.  The
analysis was specifically designed to detect regions of excess and/or
cavities in the x-ray surface brightness which were coincident with
the radio plasma. This was done by comparing the observed cluster to a
smooth, isothermal model of the emission created within the {\tt
IRAF/STSDAS} environment. Using the task {\tt ELLIPSE}, a fit to the
ellipticity, position angle and intensity of the emission was
performed. Since the goal was to create an idealized model, free of
structural quirks, all obvious features and point sources in the
images were masked out and the centroid position was fixed at the
point of peak intensity. After the models were created, they were
subtracted from the original cluster images, generating residual
maps. The residual images were smoothed with a $\rm FWHM = 20''$
Gaussian and overlaid onto the radio maps as shown in Figs. 1-3.

The residual maps provide strong observational evidence for a
connection between the hot, x-ray gas and the radio plasma. A133 and
A2626 both show significant enhancements in the x-ray emission near
the radio jets and A2052 shows a large ridge of excess x-ray emission
surrounding much of the steep spectrum radio emission. As described in
\S 1, such observational features may imply that the jet has
interacted with the cluster gas.

Another method of determining whether the x-ray emitting gas and radio
plasma are indeed interacting is through examining the shift in the
isophotal fit parameters of the x-ray emission (\cite{owen98}).  We
chose to focus our analysis on the shift in the emission weighted
centroid since this has been shown to be a sensitive measure of
substructure (Mohr \etal 1995). Using the {\tt IRAF} task {\tt
ELLIPSE}, an isophotal analysis of the x-ray surface brightness was
again employed, fitting for the centroid position, as well as
intensity, ellipticity and positional angle of the emission as a
function of radius. The analysis was performed on the raw count images
and the step size in semi-major axis, $a$ was set to twice the FWHM of
the HRI point response. As a check of the {\tt ELLIPSE} results, the
centroid shift was also determined using a moment analysis based on
the method of Buote \& Canizares (1994).  The code we used was adapted
to perform the analysis in concentric circular annuli, thus increasing
the sensitivity to shifts in high surface brightness regions such as
the cores of cooling flows. Both methods produced nearly identical
results for the centroid shifts, assuring the robustness of this test.
Fig. 4 shows the centroid position as a function of radius for the
inner regions of all three clusters.  The profiles were smoothed with
a boxcar average in order to enhance the shifts in the centroid
position for the three sources (A133 and A2052 in particular) at
angular scales ($\sim 20-30''$) corresponding roughly to the distance
of the steep spectrum radio emission from the x-ray center.

\section{Emission mechanism}
The previous analysis indicates that x-ray substructure is present in
the three clusters examined.  The model proposed here interprets the
modification of the x-ray emission to be a result of an interaction
between the radio plasma and the intracluster medium. However, for the
two cases where x-ray excesses were observed to be coincident with the
radio emission (A133 and A2626) the possibility of a non-thermal
origin for the emission cannot be entirely ruled out.  Non-thermal
x-ray emission could be produced by inverse Compton (IC) scattering of
the 3K microwave background radiation off the relativistic electrons
in the radio plasma (Harris \& Grindlay 1979).  This process would
manifest itself in spatially coincident x-ray and radio structures as
seen observationally. Ideally, one would like to examine the spectral
signatures of the regions in order to distinguish between a power law
or thermal spectrum. Unfortunately, the small angular scales of these
structures precludes using the PSPC data for A133. The HRI data has
adequate spatial resolution but negligible spectral resolution and
therefore is also incapable of differentiating between the two
emission models.  Upcoming x-ray missions (\eg AXAF, XMM) will provide
the sensitivity and resolution necessary for such a distinction.

As an alternate approach to addressing this issue, the excess emission
was assumed to be due to IC scattering and the implications for the
underlying magnetic field were examined. Under the IC assumption, one
can calculate the magnetic field necessary to create an x-ray excess
using the expression from Harris \& Grindlay (1979),

\begin{equation}
B^{(\alpha+1)} = \frac{(5.05 \times 10^4)C(\alpha)G(\alpha)
(1+z)^{(\alpha+3)}S_r\nu^{\alpha}_r}
{10^{47}S_X\nu^{\alpha}_X}
\end{equation}

\noindent where $B$ is the magnetic field and $C(\alpha)$ is a function 
of the spectral index of the source and adopted to be $C(\alpha) =
1.15 \times 10^{31}$ for $0.5 < \alpha < 2.0$ (Harris \& Grindlay
1979) .  $G(\alpha)$ is a correction factor for the energy
distribution of the microwave background radiation with a value on the
order of unity, $S_r$ (ergs $\rm cm^{-2}$ $\rm s^{-1} Hz ^{-1})$ is
the observed radio flux density at the observed frequency $\nu_r$ and
$S_X$ (ergs $\rm cm^{-2}$ $\rm s^{-1} Hz ^{-1})$ is the observed x-ray
flux density at the observed x-ray frequency, $\nu_X$. Since $S_X$ and
$S_r$ are both measured quantities, one can estimate the magnetic
field for the two clusters in a rather straightforward manner.  The
resultant fields, ${\bf B}_{\rm IC}\sim 0.02$ and $\sim 0.3 $ $\mu
\rm G$ for A133 and A2626, respectively, are listed in Table 3 along
with the flux densities and spectral indicies used for both cases (see
below).  For comparison, the equipartition fields in these regions,
${\bf B}_{\rm eq}$ were calculated over the frequency range of
0.29-4.885 GHz and 0.38-4.885 GHz for A133 and A2626 respectively. A
volume filling factor of $\phi = 1$ was assumed along with equal
contributions of proton and electron energies (k=1).  These values are
relatively uncertain due to the required assumptions about the
geometry of the system (in this case taken to be cylindrical),
however, it serves as a framework for comparison with our IC derived
magnetic field results.

As listed in Table 3, ${\bf B}_{\rm eq} < {\bf B}_{\rm IC}$, yet there
are several caveats which hint at a contribution from a non-thermal
source for the x-ray emission. Given a 1 keV photon detected in the
x-ray and $B \sim 2 \mu \rm G$, a microwave background photon near the
peak of the Planck function ($\nu \approx 1.6 \times 10^{11} \rm Hz$)
would need to be scattered by synchrotron producing electrons at about
10 MHz.  Unfortunately, the overall spectral index measurements for
both sources were derived from higher frequency observations. Since
the magnetic field calculation is a strong function of $\alpha$, the
very high spectral indices would greatly affect the resultant
calculations by predicting an enormous population of low energy
electrons. In reality, the spectra of the radio sources are likely to
flatten out at low frequencies.  We therefore fit the flux densities
of the two sources using only the lowest frequency data available
($\sim 30-100$ MHz for these clusters) and determined $\alpha = 0.9$
and $\alpha = 1.33$ for A133 and A2626 respectively over
this spectral region. Low frequency observations are far more
difficult to obtain and lack the resolution acheivable in shorter
wavelength data. Consequently, the spectral indicies we calculate may
be contaminated by flatter spectra components of the source and
therefore lead to an underestimate the IC determined magnetic field.
 
This calculation is also biased by the assumption that all the
observed x-ray excess is due to IC scattering. Since these sources
both reside at the centers of clusters of galaxies, there is a
significant thermal component pervading these regions. We have attempted
to remove this component (see \S 5), however, it is likely that a
contribution from the cluster ICM may still be present.  This
contamination acts in the same sense as the effect above. Since the
magnetic field and x-ray flux are inversely related, an artificially
high $S_X$ would lead to an erroneously low ${\bf B}_{\rm IC}$.

\section{Modeling the Jet-Cooling Flow Interaction}

An alternative explanation for the observed spatial distribution of
the x-ray and radio emission is the hydrodynamical interaction of jet
plasma and the ambient medium which can result in x-ray surface
brightness deficits and excesses (e.g., Carilli, Perley, \& Harris
1994; Clarke, Harris, \& Carilli 1997). In order to investigate these
effects further, we have carried out two new simulations of perturbed
3D jets propagating into a cooling flow atmosphere (see e.g., Loken et
al.~1993 for similar simulations in 2D).

The simulations were performed using ZEUS-3D (Stone \& Norman 1992), a
three-dimensional MHD code maintained by the Laboratory for
Computational Astrophysics at NCSA . The computational volume is a
quadrant from a sphere (2.0 kpc $\le R \le$ 25.0 kpc, and $5^0 \le
\theta, \phi \le 85^0$) with 460 uniform zones in the radial direction
and 110 zones in each of the $\theta$ and $\phi$ directions,
concentrated along the jet axis. In both simulations, the jet is
injected at the center of the inner boundary (R=2kpc,
$\theta=\phi=45^0$) with a radius of 0.25 kpc.  The jet density is ten
times less than that of the cooling flow atmosphere (at the inlet) and
it is pressure-matched with the ambient medium.  The grid was
initialized with the density, temperature and velocity profile
corresponding to a spherically-symmetric, steady-state cooling flow
obtained by integrating the standard, 1D, steady-state cooling flow
equations (e.g., White \& Sarazin 1987). These equations were
integrated inward from the cooling radius ($r_{cool}$) assuming a
temperature $T=4 \times 10^7$K (3.4 keV) and a mass inflow rate,
$\dot{M} = 100 M_\odot yr^{-1}$, at $r_{cool}$.  The resultant
profiles for the atmosphere are shown in Fig.~\ref{coolflow}. Note
that these simulations make use of simplified initial conditions and
limited physics and thus are not intended to model a specific
source. Nevertheless, the results exhibit features which aid in the
interpretation of our data.

In our first simulation, a jet with internal Mach number M$_j$=5
propagates onto the grid and disrupts at R$\sim$ 5kpc due to the
applied perturbation.  The density distribution along the jet
mid-plane is shown in Fig.~\ref{jetdens} at the final epoch
(t=1.1$\times 10^7$ years). At this point, the jet has completely
disrupted and its advance has essentially stopped.  A large, turbulent
lobe of processed jet material fills much of the central region of the
image ($7 r_j \le R \le 15 r_j; \theta \sim 45^o$).  In order to
construct an x-ray residual map akin to our observational results, we
calculated the Raymond-Smith emission in the (0.1-2.4) keV bandpass
for every zone in the simulation at the final epoch after subtracting
the emission due to the initial cooling flow atmosphere.  This
residual x-ray emission was then summed along the line of sight to
produce Fig.~\ref{xray_af}, assuming the jet to be at the same
redshift as A133. Notice that the bow-shock surrounding the jet is
clearly delineated in this image and that enhanced emission is
apparent from much of the cocoon region. Interestingly, there is a
hole, or deficit, in the emission along lines of sight passing through
the disrupted jet lobe.  A more quantitative interpretation of the
x-ray deficits and excesses can be obtained from the slices across the
x-ray image which are shown in Fig.~\ref{xrayslices}. Each slice shows
the x-ray surface brightness profiles at the final epoch (solid line)
and from the initial, steady-state cooling flow atmosphere (dotted
line). The first slice is taken just above the jet and thus shows no
difference between the original and final epochs. Panel (b) shows an
x-ray excess from the leading edge of the bow shock. The remaining
cuts again show excesses from the bow-shock region but deficits along
the jet axis are also apparent.

The second simulation utilizes the same set-up and parameters as the
first except that the jet Mach number in substantially increased
(M$_j$=20) and the inflow at the inlet is turned off 2.2 Myrs before
the final epoch. Fig.~\ref{dens_ak} shows density contours for the jet
just before it is turned off and at the final epoch. Despite the fact
that there is no new momentum or kinetic energy injected after the
epoch shown in the top panel, the jet and its bow-shock continue to
propagate outward under their residual momentum. At the final epoch
shown, the outward motion has almost completely ceased and the jet
remnant (the material within the contact discontinuity delineated by
the closely-spaced contours) is collapsing inward on itself since
there is no energy input to counteract the high-pressure of the
surrounding cocoon.  The residual x-ray image for the final epoch is
shown in Fig.~\ref{xray_ak}. Clearly, turning-off the jet has resulted
in obvious differences from the case shown in
Fig.~\ref{xray_af}. Though this jet is initially more powerful, there
is currently no source of kinetic energy to power the strong, multiple
bow-shocks so evident in Fig.~\ref{xray_af}. In addition, the remnant
of the jet itself is denser than it would be if the jet had not been
turned off, making it difficult to trace the jet via the x-ray deficit
as in Fig.~\ref{xray_af}. As before, there is still enhanced emission
visible from the edges of the bow-shock and some deficits inside the
cocoon region (primarily near the inlet).

No magnetic fields or relativistic particles were included in the
numerical simulations so we cannot make detailed predictions of the
resultant radio morphologies. Nonetheless, it is likely that radio
emission is limited to the regions containing jet plasma (i.e. within
the contact discontinuity separating cocoon material from shocked
ambient) unless there has been substantial diffusion. In the case of
the M$_j$=5 simulation, this would imply that the radio emission is
spatially coincident with x-ray deficits and surrounded by regions of
x-ray excess.

\section{Discussion}
Observationally, we see rather compelling evidence of an interaction
between the hot, x-ray emitting intracluster medium and the radio
plasma from the central, steep spectrum radio sources in three cooling
flow clusters. This evidence is apparent in the presence of
substructure as measured through the x-ray emission weighted centroid
shifts and the spatial coincidence (or anti-coincidence in the case of
A2052) of the x-ray excesses and the steep spectrum component of the
radio sources. Although no attempt has been made to specifically model
any of the three sources, the numerical jet/lobe simulations reproduce
a number of the observational features. In particular, x-ray excesses
are visible from regions where the gas has been compressed and heated
by the passage of the jet bow shock.

The slices of the projected surface brightness distribution across the
bow shock of the jet (Fig.\ref{xrayslices}) exhibit x-ray excesses on
the order of those detected observationally for both A133 ($\sim 1.5
\times 10^{-15}$ $\rm ergs/s/cm^{2}/arcsec^{2}$) and A2626 ($\sim 4.5
\times 10^{-16}$ $\rm ergs/s/cm^{2}/arcsec^{2}$).  The excess emission
associated with A2052 is slightly higher than the other two ($\sim 4.5
\times 10^{-15}$ $\rm ergs/s/cm^{2}/arcsec^{2}$), yet still within an
order of magnitude.  One can imagine a scenario where the bow shock of
the propagating jet compresses and heats the ambient medium as it
passes through the cluster core, before the jet eventually
disrupts. The increased density, $n$, and temperature, $T$, caused by
the shock boosts the x-ray emissivity, $\epsilon$, from that region
($\epsilon \propto n^2 T^{1/2}$) which is manifested in an increase in
the surface brightness.  A similar situation is believed to exist
along the lobe edges in Cygnus-A, where shocked material has increased
in density and emissivity (Carilli, Perley \& Harris 1994).

The unusual morphology of A133 and A2626 suggest that the jets have
significantly disrupted and the shocked, diffuse plasma has spilled
over, perhaps abutting the compressed emission from the original bow
shock.  This scenario is supported by the very steep spectrum of the
diffuse structure, which suggests that the extended region of the
source is surrounded and confined by the dense ICM preventing the loss
of energy through expansion and leading to the steepening of the
source's spectrum. From simple pressure arguments, one can show that
the radio sources in A133 and A2626 are indeed under pressured.  Using
the densities determined through x-ray deprojection and a central
temperature of about $2 \times 10^7$K ($\sim 2$ keV) results in
thermal pressures ($P_g = nkT$) which are about 2 orders of magnitude
higher ($P_g \sim 10^{-11}$) than the magnetic pressures ($P_m \sim
10^{-13}$) computed from equipartition (see Table 3).

The observational characteristics of A2052 are somewhat more
complicated.  This radio source, 3C317, exhibits a very bright core
and diffuse outer halo, but no jets resolved at the $0.3''$
level. Zhao \etal (1993) suggest that the radio source was disrupted
at a very early stage due to interaction with the cooling flow
environment at the core of the cluster.  The x-ray excess seen in the
HRI data surrounds roughly half of the diffuse halo. This emission is
again presumably a signature of material compressed by the initial bow
shock and/or from inflowing material which is compressed upon contact
with the hot, diffuse gas from the disrupted jet in the central
regions.  For this cluster, the equipartition field ($20 \mu$G),
yields comparable thermal and minimum energy pressures for the diffuse
halo (\cite{zhao93}).

Along with x-ray excesses, the numerical simulations also produce
x-ray deficits along lines of sight passing through the region where
the disrupted jet has displaced the original ICM.  However, strong
x-ray deficits (as in Fig.~\ref{xray_af}) are not seen in any of our
observed sources. It is possible that the viewing geometry could hide
these deficits but even when the M$_j$=5 jet is viewed head-on, the
x-ray residual map shows a significant hole along the jet axis
surrounded by rings of x-ray excesses. This is perhaps not surprising
as in this case there will be lines of sight with even longer path
lengths through the disrupted jet material than in the projection of
Fig.~\ref{xray_af}. Another possible explanation for this discrepancy
between the M$_j$=5 simulation and our observed sources is that no
distinction has been made between thermal and non-thermal plasma in
the code. This essentially results in a synthetic x-ray image with a
filling factor of 1 for the relativistic plasma, thereby maximizing
the depth of the x-ray deficits. A smaller assumed filling factor
would reduce the spatial extent and depth of the deficits, perhaps
removing them entirely. However, we note that, similar to
Fig.~\ref{xray_af}, there are sources where deficits appear to be
coincident with radio emission. For example, the outer radio lobes of
3C84 in NGC 1275 correspond to minima in the x-ray surface brightness
suggesting that the relativistic jet plasma has displaced all the
thermal gas (Bohringer et al.~1993). Furthermore, the factor of 2
difference in x-ray surface brightness between the radio lobes and the
surrounding regions in NGC 1275 is similar to that seen in
Fig.~\ref{xrayslices}.  Lastly, the lack of observed x-ray deficits
may be an artifact of the necessarily imprecise subtraction of the
spherical cooling flow atmosphere. Only in the case of the simulation
do we know the exact model for the underlying cluster emission to be
used in creating the residual map.

The jet which has been turned off (Figs.~\ref{dens_ak} and
\ref{xray_ak}) perhaps most closely resembles our observed
sources. This simulation still exhibits x-ray excesses along the bow
shock (and therefore surrounding the radio emission) but the deficits
along the jet are much less pronounced. Of course, some of the
differences between our simulated and observed sources may be due to
our assumption of an initially symmetric cooling flow atmosphere. In
at least the case of A133, it is possible that a dense knot of gas to
the north of the cluster center is responsible for both
deflecting/disrupting the jet and for creating the x-ray excess. In
such a case, the jet may not be able to displace much of the ambient
medium so no deficit would be apparent.

We compare the energies associated with the jets and the cooling flow
atmosphere in our simulations. The total X-ray luminosity of the
initial atmosphere on the computational grid with 2 kpc
$\leq$R$\leq$10 kpc is $3.2\times10^{43}$ erg/s (this doubles for the
region within 20 kpc). This is virtually identical to the kinetic
luminosity of the M$_j$=5 jet ($3.5\times10^{43}$ erg/s) but
substantially less than the kinetic luminosity of $2.2\times10^{45}$
erg/s for the M$_j$=20 jet (before it is turned-off). In both cases,
the cooling time in the atmosphere (initially $\sim 1.6 \times 10^8$
yrs at the inner boundary) is much longer than the total time for
which the simulation was run. This implies that, while it is on, the
M$_j$=20 jet is a net source of energy (for the region within
$\sim10-20$ kpc) and that the M$_j$=5 jet pumps roughly as much energy
into the surrounding medium as the cooling flow loses via thermal
emission.

These results on the energetics of the sources are consistent with the
recent interpretation of the interplay between the cooling flow and
central radio source in M87 (Owen, Eilek \& Kassim, 1999). Under this
scenario, the radio source imparts a significant amount of energy into
the core region of the cluster during its very active stage,
disrupting the cooling flow and even developing a central ``heating
flow.'' As the radio source ages, it has less of an impact
energetically on the surrounding cluster environment and the cooling
flow is able to re-establishes itself.  The x-ray surface brightness
irregularities in the aged, steep spectrum sources presented here may
be evidence of that energy transfer between the radio source and
cluster environment at earlier epochs of the radio source lifetimes.

Adopting the secondary hypothesis that the x-ray excesses are
non-thermal in origin, estimates of the magnetic fields in the radio
sources can be determined through an assumption of the inverse Compton
scattering processes.  This analysis yields ${\bf B}_{IC}$ values of
$0.03$ $\mu G$ and $0.27$ $\mu G$ for A133 and A2626 respectively.
These values are considerably lower than the equipartition results,
implying that it is unlikely the excess x-ray emission is
predominantly non-thermal in origin. However, the striking projected
spatial coincidence between the x-ray excesses and steep spectra radio
sources, along with the caveats stated in \S 6 make it difficult to
entirely eliminate an IC contribution. The implication of a possible
IC detection is quite interesting in its own right. Although a
significant body of work has focused on the models and energetics of
radio sources, definitive results are often plagued by assumptions
required for uncertain quantities such as the magnetic fields. Inverse
Compton scattering provides a much desired, direct measure of the
magnetic fields within extragalactic radio sources. To date,
observations of this phenomena have only been claimed in a handful of
cases (\eg Feigelson \etal 1995; Kaneda \etal 1995; Bagchi, Pislar \&
Lima Neto 1998) so additional confirmed cases of IC scattering would
certainly be of interest. Currently the data does not allow a clear
distinction to be made between the thermal and non-thermal
contributions to the excess x-ray emission.  However, with the spatial
and spectroscopic capabilities of AXAF and XMM, it should be possible
to accurately determine the relative significance of each model.

\acknowledgements
We would like to thank Q.D. Wang for supplying us with the original
form of the moment analysis code and J.-H. Zhao for contributing the
3C317 VLA data. This work was supported in part by the NASA ROSAT
program and NASA grant NAGW-3152 to J.O.B. and F.N.O. and NSF grant
AST-9896039 to J.O.B.

\clearpage

\begin{deluxetable}{clcccccc}
\footnotesize
\tablecaption{Radio Observational Log\label{tbl-2}}
\tablehead{
\colhead{Cluster} & \colhead{Observation date}   & \colhead{Array}   & 
\colhead{Exposure time} & \colhead{$\rm S_{rms}$} &
\colhead{Frequency} & \colhead{Bandwidth}  & \colhead{S\tablenotemark{a}} \nl
\colhead{Abell/IAU (1950)} & \colhead{ }   & \colhead{ }   & 
\colhead{ } & \colhead{($\mu$Jy)}   &
\colhead{(GHz)} & \colhead{(MHz)}  & \colhead{(mJy)} }

\startdata
A0133 (0100-221A) & 1993 May 4   & B & 21:20 & 104 & 1.4 & 50  & 138  \nl
A0133 (0100-221A) & 1993 July 15 & C & 31:40 & 104 & 1.4 & 50  & 138  \nl
A2626 (2333+208)  & 1993 July 15 & C & 34:00 & 52 & 1.4 & 3   & 55   \nl
\enddata

\tablenotetext{a}{Total flux densities from Ledlow \& Owen (1995)}
\end{deluxetable}


\begin{deluxetable}{cllllc}
\footnotesize
\tablecaption{X-ray Observational Log\label{tbl-1}}
\tablewidth{475pt}
\tablehead{
\colhead{Cluster} & \colhead{RA(1950)}   & \colhead{DEC(1950)}   & 
\colhead{Sequence Number} & \colhead{Live time}  & \colhead{$z$}} 
\startdata
A0133 & 01:00:09.3 &-22:02:55 & rh800428a01 & 9900   & 0.0573  \nl
A0133 & \nodata    & \nodata  & rh800428a02 & 12425  & \nodata \nl
A0133 & \nodata    & \nodata  & rh800428n00 & 9050   & \nodata \nl
A0133 & \nodata    & \nodata  & rp800319    & 15380  & \nodata \nl
A2052 & 15:14:15.7 & 07:12:11 & rh800223a01 & 808    & 0.0348  \nl
A2052 & \nodata    & \nodata  & rh800223a02 & 3835   & \nodata \nl
A2052 & \nodata    & \nodata  & rh800223n00 & 4385   & \nodata \nl
A2052 & \nodata    & \nodata  & rp800275n00 & 6008   & \nodata \nl
A2626 & 23:34:00.2 & 20:52:24 & rh800430a01 & 21468  & 0.0604  \nl
A2626 & \nodata    & \nodata  & rh800430n00 & 6168   & \nodata \nl
 
\enddata
\end{deluxetable}

 
\begin{deluxetable}{cllllc}
\footnotesize
\tablecaption{Magnetic Field Estimates\label{tbl-3}}
\tablewidth{475pt}
\tablehead{
\colhead{Cluster} & \colhead{$\alpha$}  & \colhead{$\rm S_X$}          & 
\colhead{$\rm S_r$}   & \colhead{{\bf $\rm B_{IC}$}}  & \colhead{{\bf $\rm B_{eq}$}} \nl
\colhead{ }       & \colhead{ } & \colhead{ ergs $ \rm s^{-1}$ $\rm  cm^{-2}$} & 
\colhead{ergs $ \rm s^{-1}$ $\rm  cm^{-2}$}   & \colhead{$\mu \rm G$}  & 
\colhead{$\mu \rm G$} }

\startdata
A0133 &  0.90\tablenotemark{a} & $4.85 \times 10^{-12}$  & $1.10 \times 10^{-1}$  & 0.02 & 2.6 \nl
A2626 &  1.33\tablenotemark{b} & $5.27 \times 10^{-14}$  & $7.53 \times 10^{-3}$  & 0.3  & 1.6 \nl
\enddata

\tablenotetext{a}{Spectral index over range 30-160 MHz.}
\tablenotetext{b}{Spectral index over range 38-178MHz}
\end{deluxetable}

\clearpage

\begin{figure}[p]
\centerline{
\psfig{figure=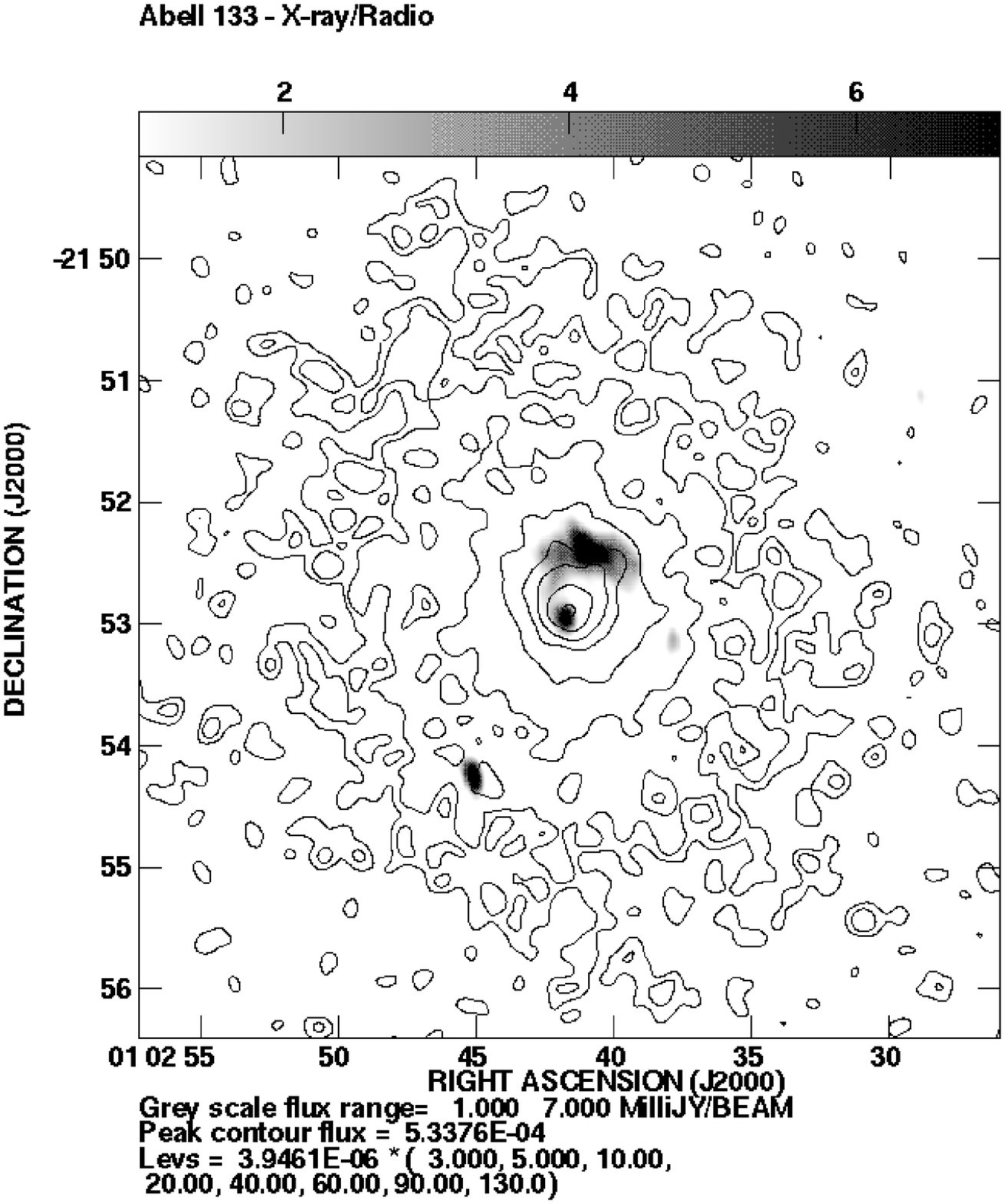,height=4.3in}
\psfig{figure=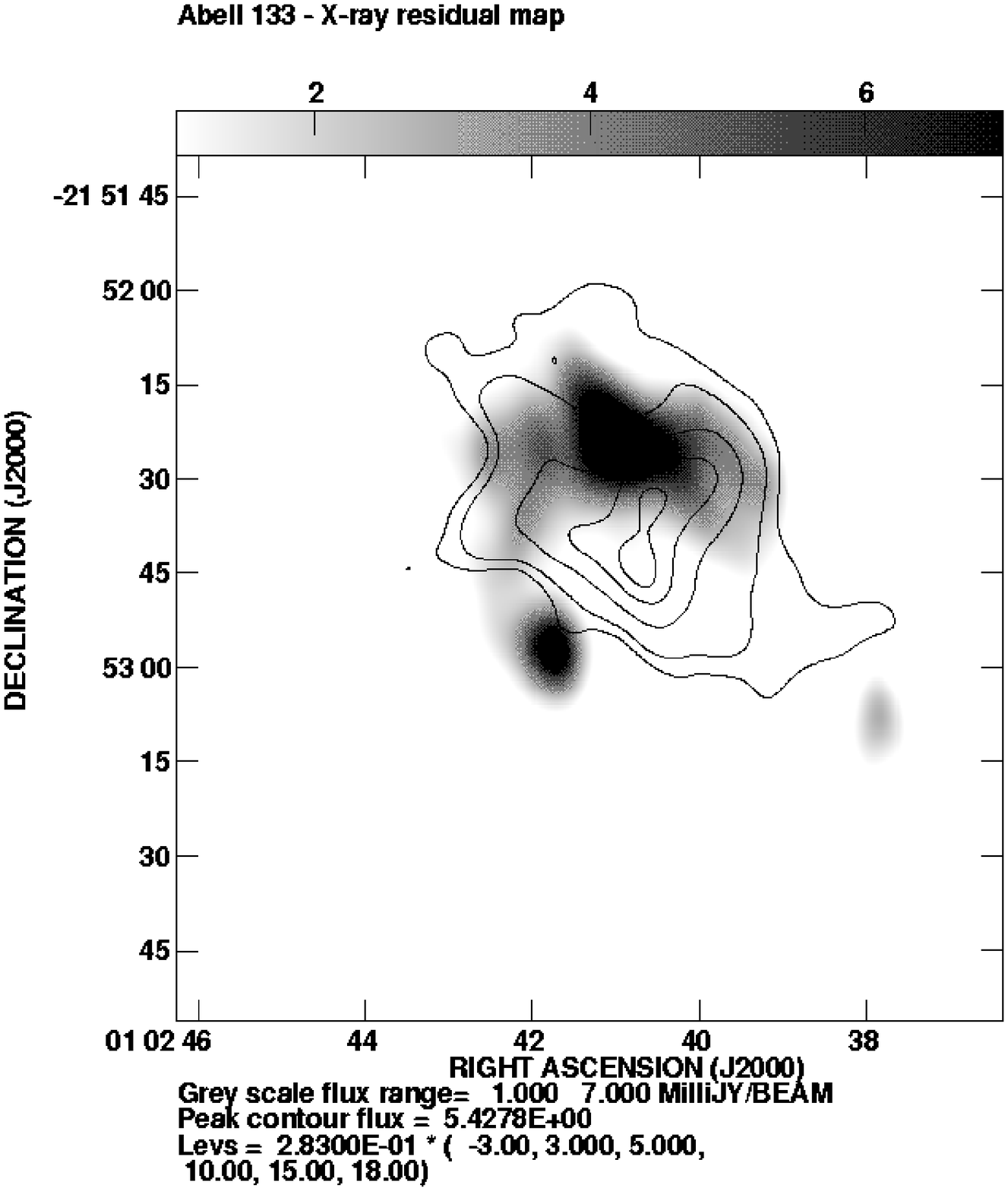,height=4.3in}}
\caption {X-ray/radio overlays for A133: panel a) shows the radio 
(greyscale) map overlaid with HRI (contours) distribution of the
cluster gas. The contour levels for the x-ray emission are (3, 5, 10,
20, 40, 60, 90, 130)$\sigma$.  Panel b) shows the x-ray
residual map smoothed with a $20''$ FWHM Gaussian after an elliptical
model has been subtracted from the overall cluster emission. The
contour levels for the residual emission are (-3, 3, 5, 10, 15,
18)$\sigma$. The linear scale at the redshift of A133, $1' \sim 60$
kpc.}
\label{A133}
\end{figure}

\begin{figure}[p]
\centerline{
\psfig{figure=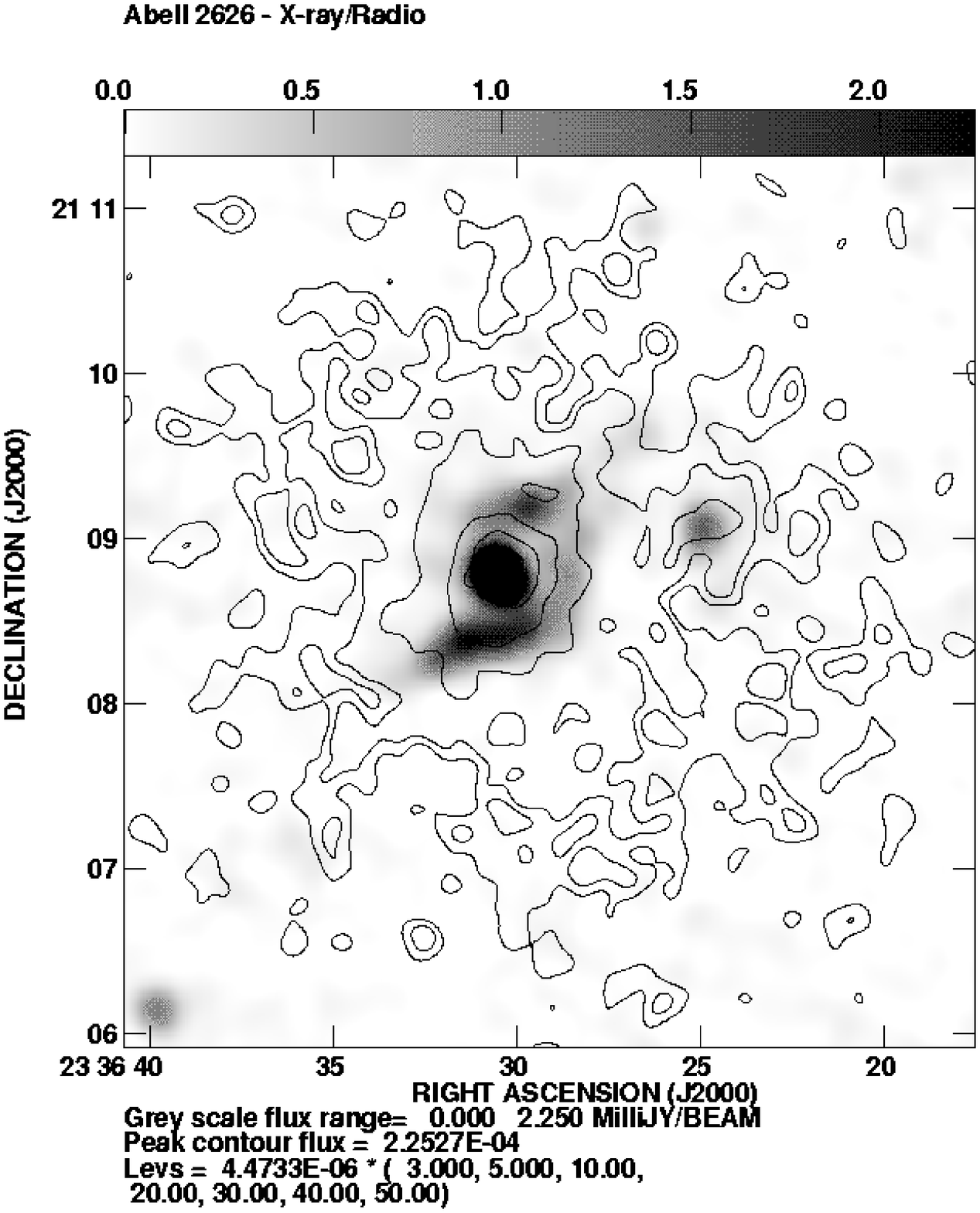,height=4.3in}
\psfig{figure=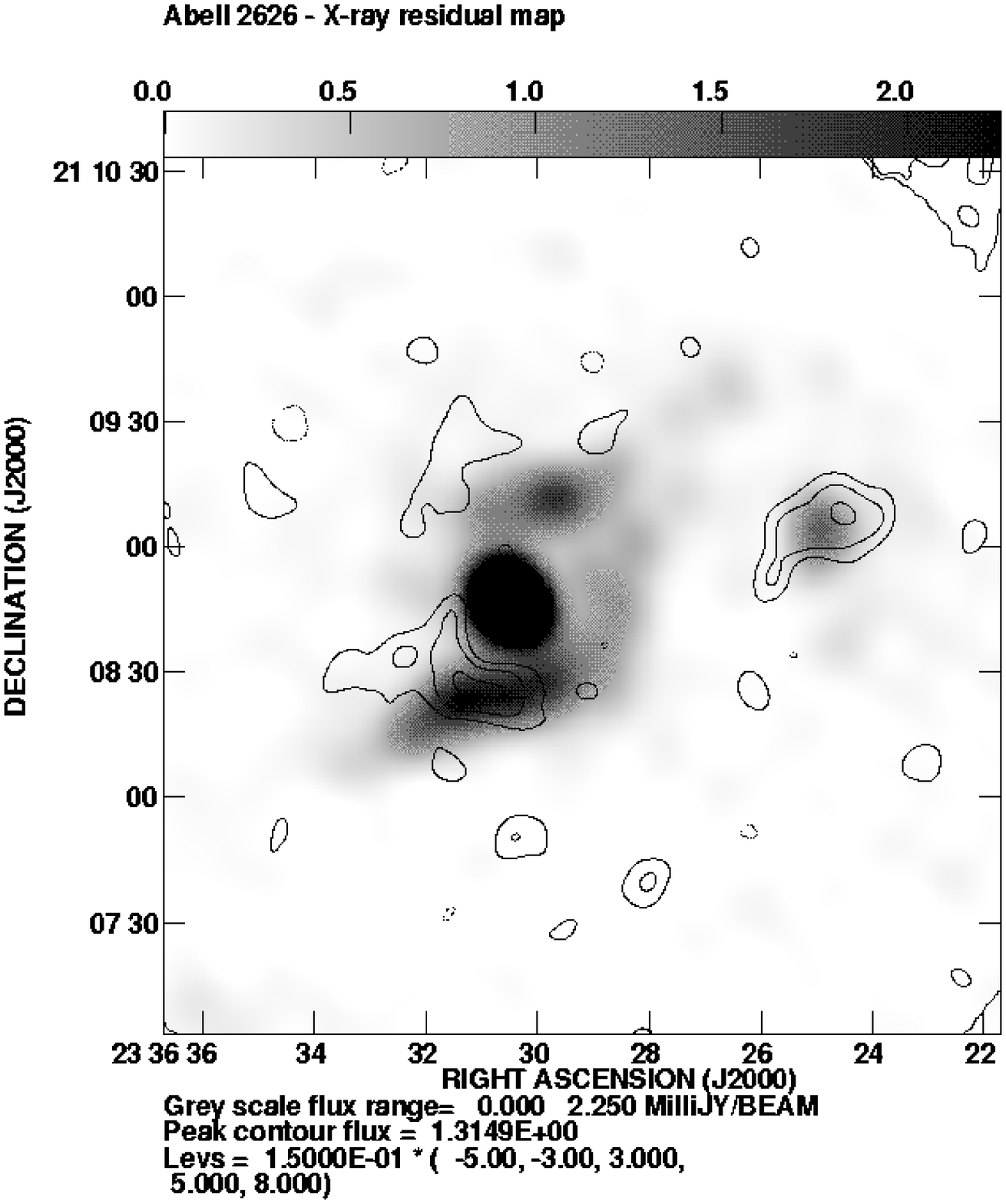,height=4.3in}}
\caption {X-ray/radio overlays for A2626: the contour
levels for the x-ray emission are (3, 5, 10, 20, 30, 40, 50)$\sigma$
and for the residual emission are (-5, 3, 5, 8)$\sigma$. The linear scale
at the redshift of A2626 is $1' \sim 65$ kpc. The spatially coincident 
x-ray/radio source at $\alpha (\rm J2000) = 23^h36^m25.1^s, \delta = 
21^{\circ}09'02''$ is the radio loud S0 galaxy IC 5337 and not examined
as part of this study.}
\label{A2626}
\end{figure}

\begin{figure}[p]
\centerline{
\psfig{figure=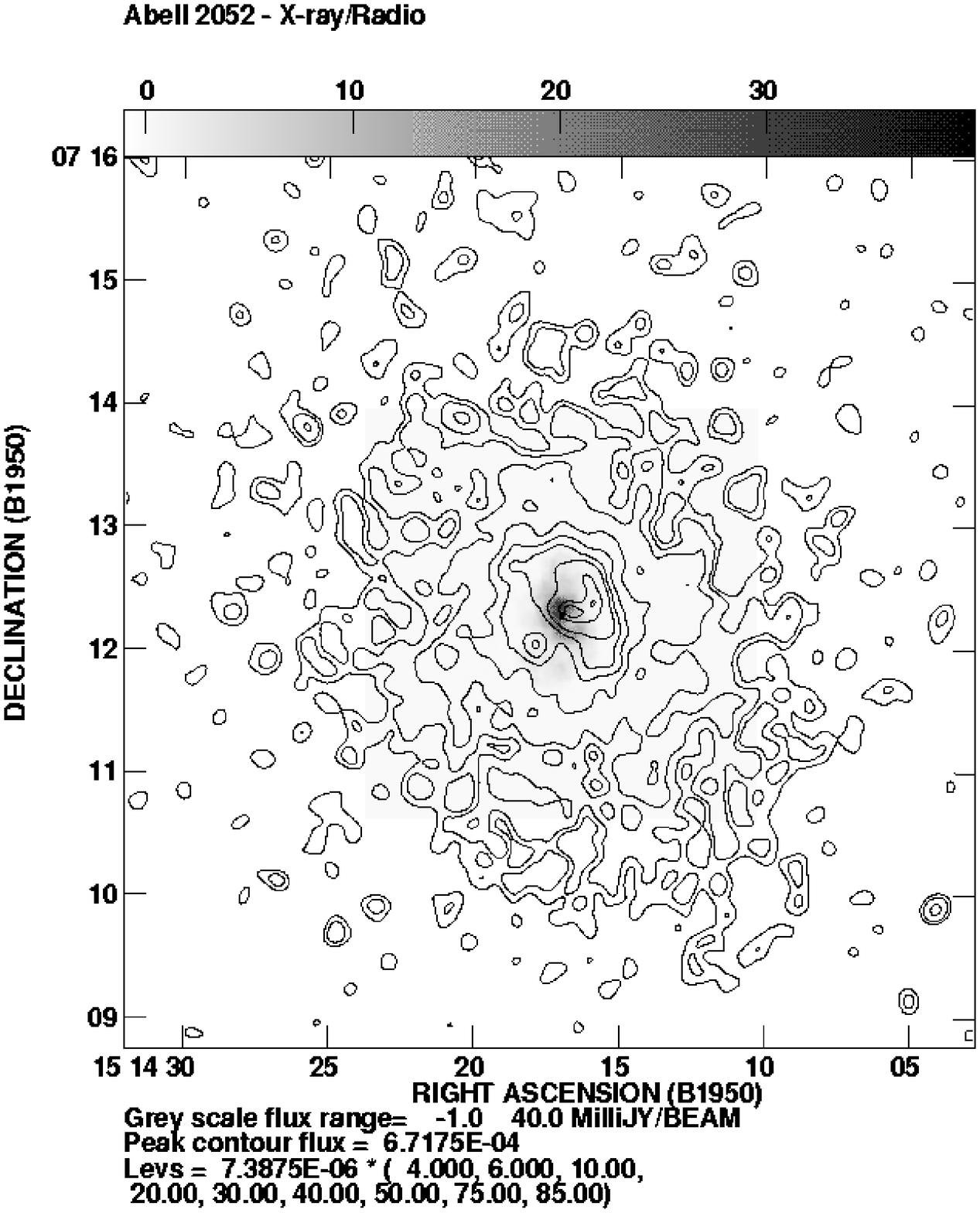,height=4.3in}
\psfig{figure=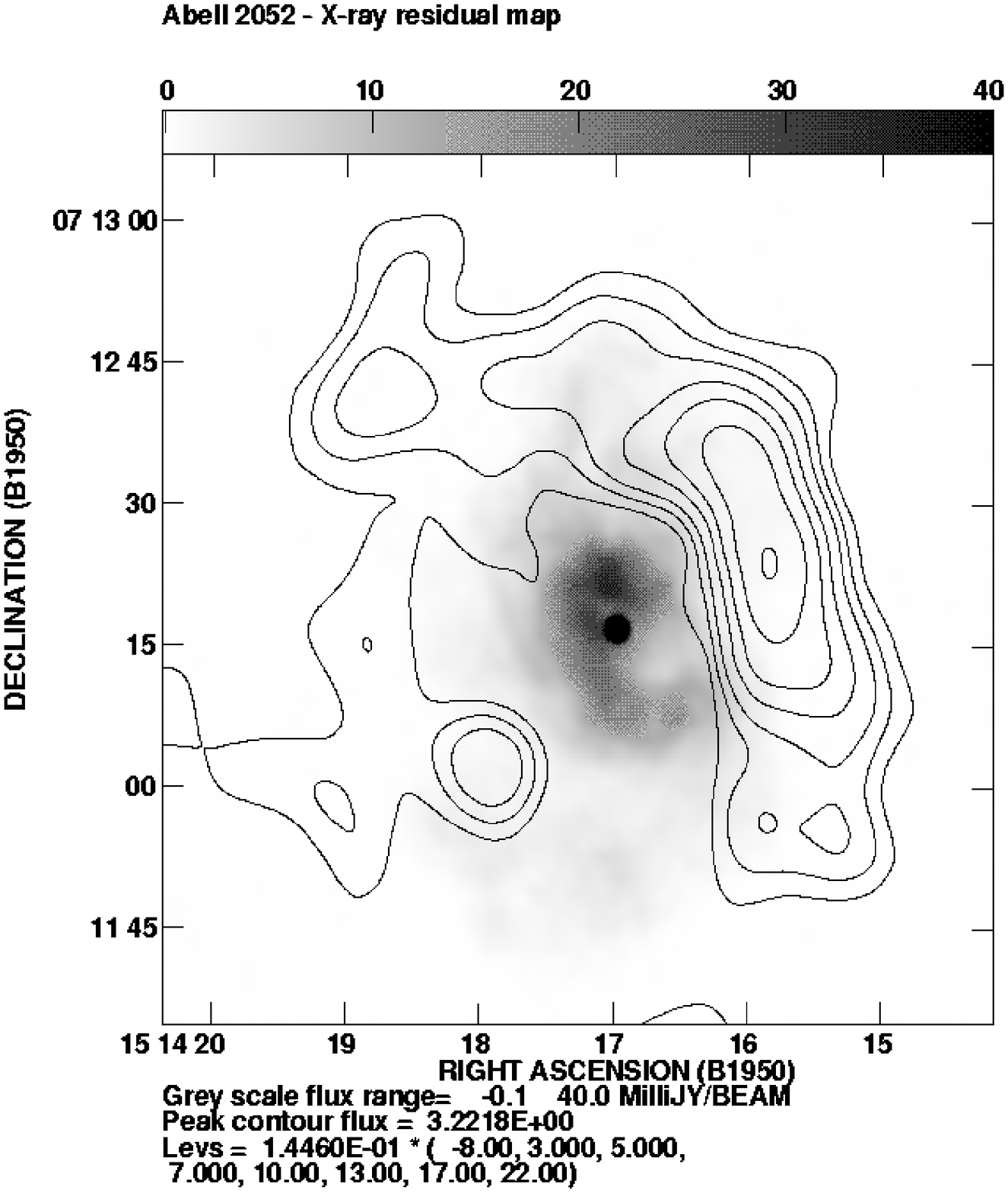,height=4.3in}}
\caption {X-ray/radio overlays for A2052: the contour
levels for the x-ray emission are (4, 6, 10, 20, 30, 40, 50, 75, 
85)$\sigma$ and for the residual emission are (-8, 3, 5, 7, 10, 13, 17,
22)$\sigma$. The linear scale at the redshift of A2052 is $1' \sim 40$ kpc.}
\label{A2052}
\end{figure}

\begin{figure}[p]
\plotone{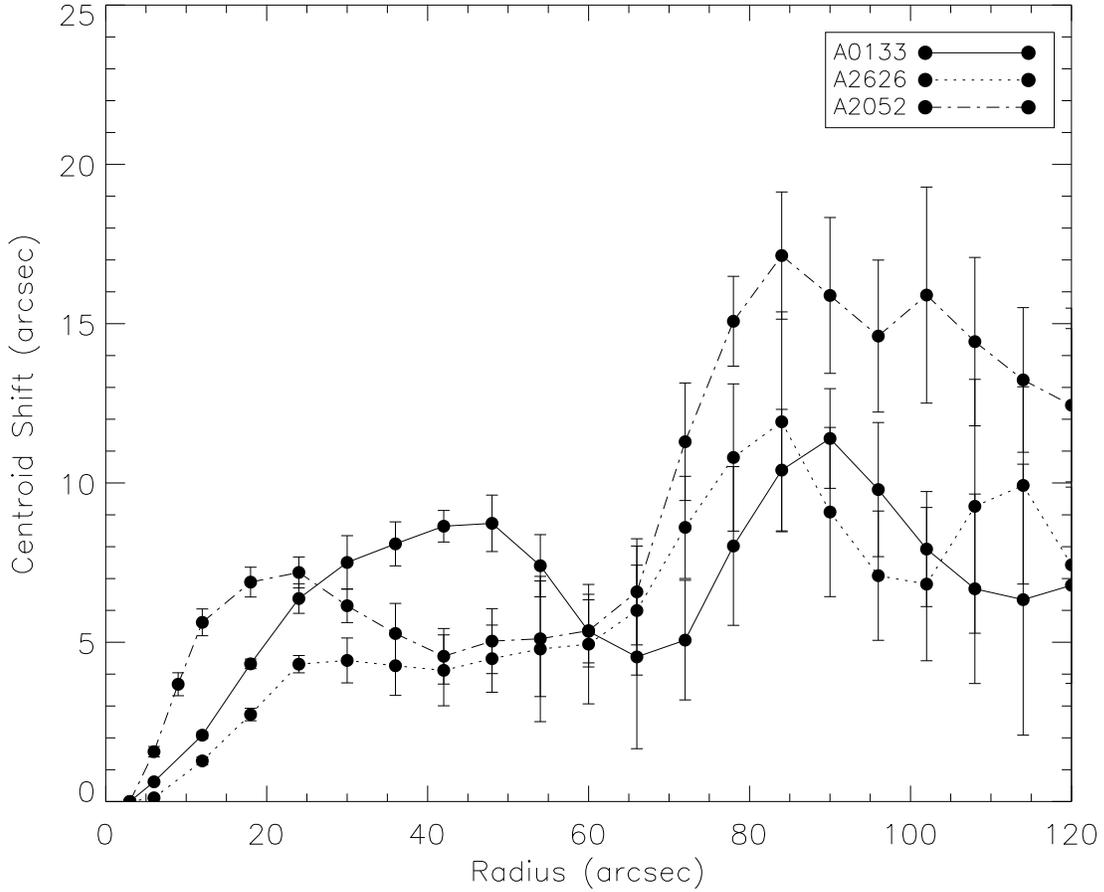}
\caption {As a measure of substructure, the shift in the
emission weighted centroid position in the core was determined from
the original count maps for each cluster. The profiles were smoothed
with a boxcar average to better display the shifts in the centroid 
position for all three sources (A133 and A2052 in particular) at 
angular scales roughly corresponding to the distance of the steep 
spectrum radio emission from the x-ray center ($\sim 20-30''$).}
\label{shift}
\end{figure}

\begin{figure}[p]
\plotone{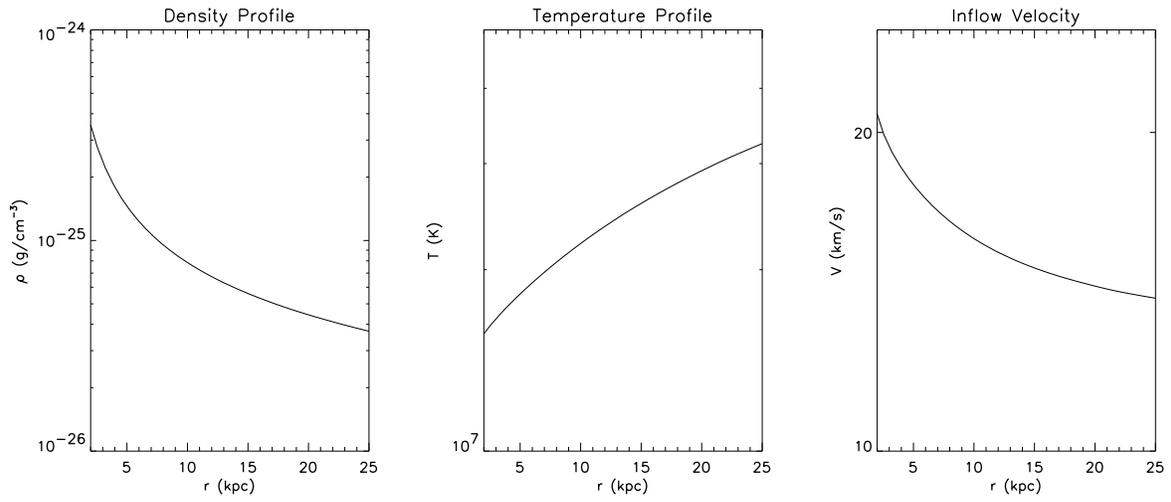}
\caption{Density, temperature and velocity profiles for the 
spherically-symmetric, steady-state cooling flow
atmosphere used to initialize the jet simulations.}
\label{coolflow}
\end{figure}

\begin{figure}[p]
\plotone{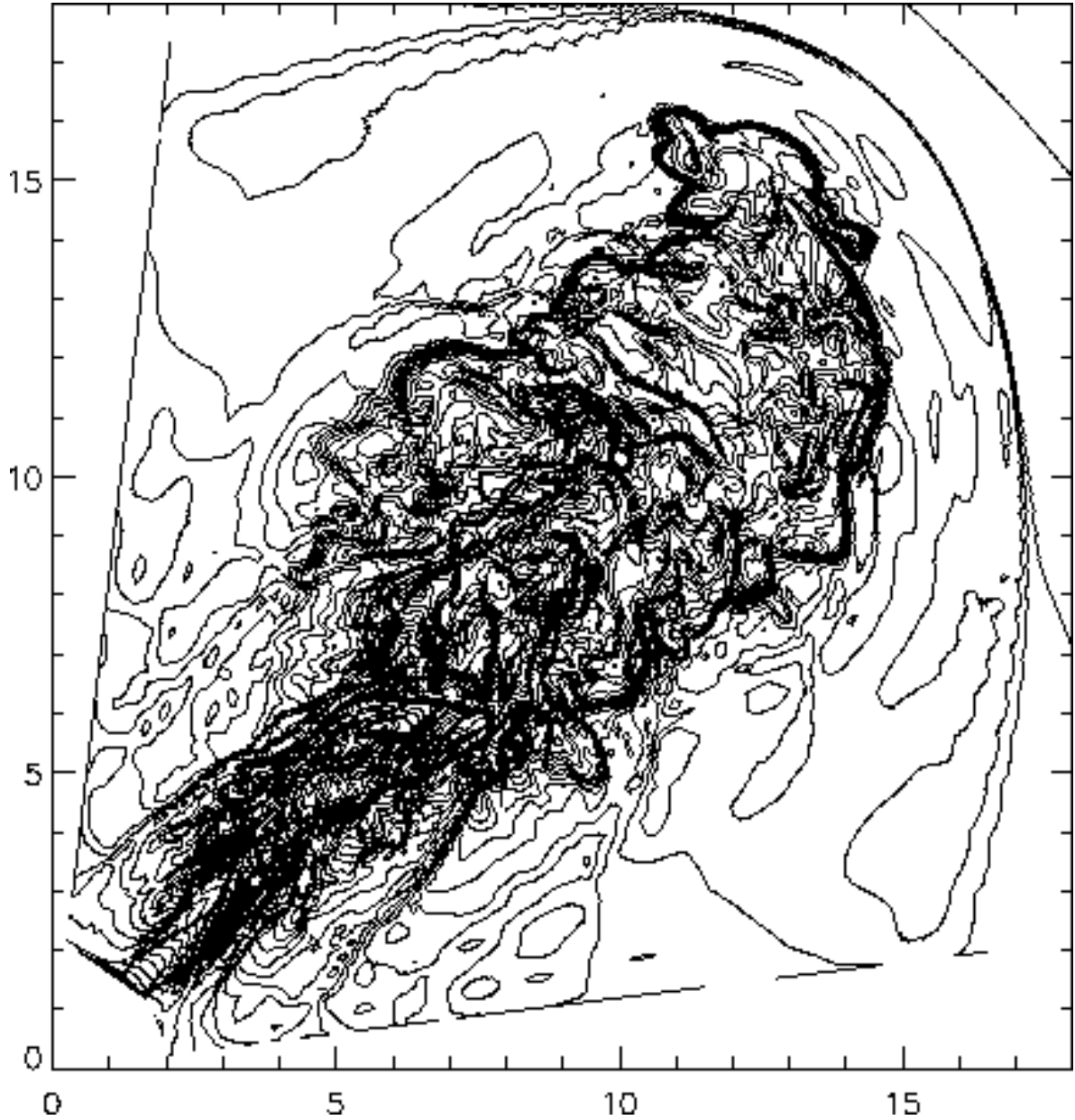}
\caption{A slice of log(density) through the mid-plane of the 
M$_j$=5 jet at the final epoch. The bow-shock has reached the
boundaries of the plotted region while the jet material is contained
within the contact discontinuity revealed by the densely-spaced
contour lines. The jet has disrupted but continues to inflate a
low-density lobe.}
\label{jetdens}
\end{figure}

\begin{figure}[p]
\plotone{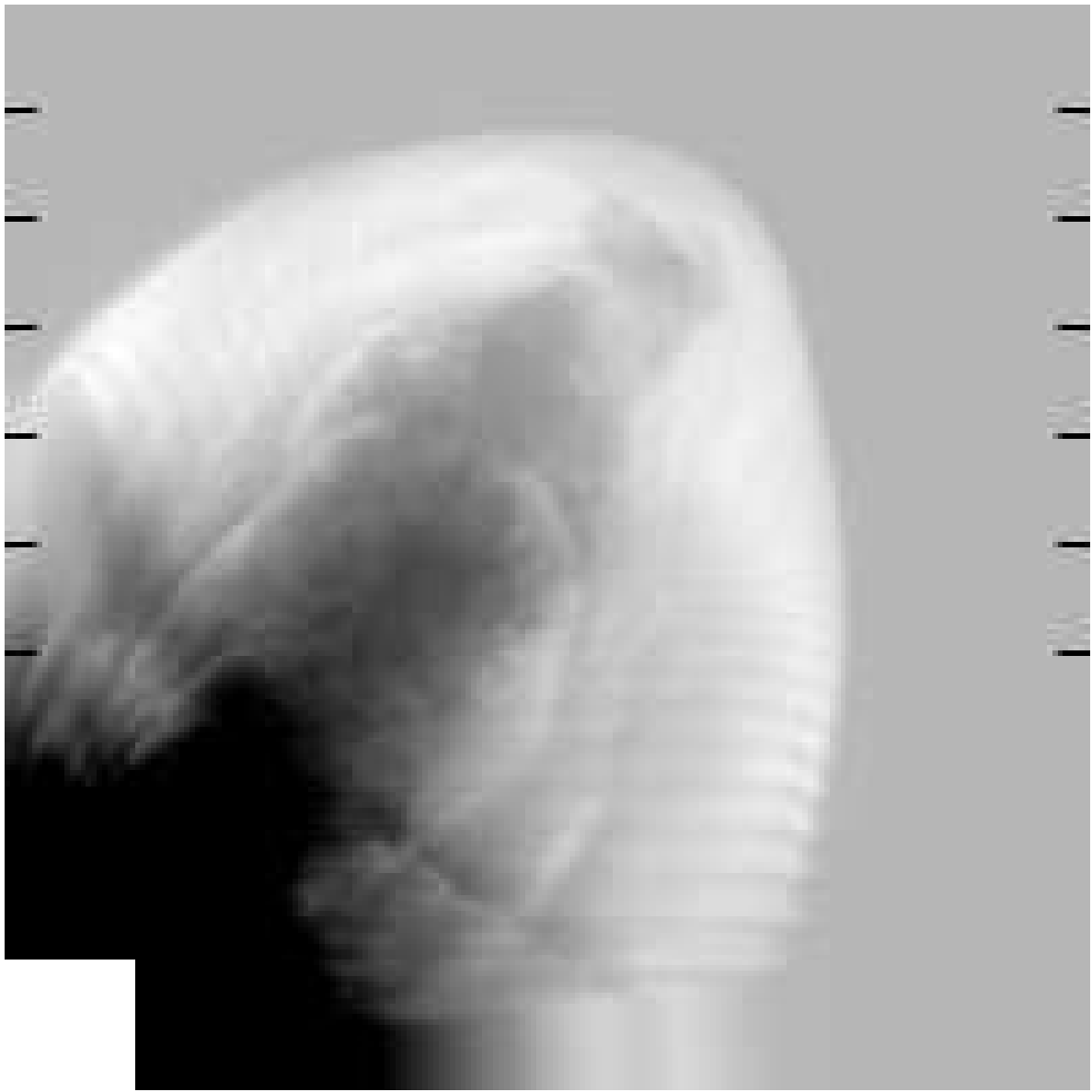}
\caption{Synthetic x-ray image of the jet at the final epoch. The initial
cooling flow atmosphere has been subtracted in order to mimic the procedure
used to create the observed residual maps. The uniformly-shaded region
along the upper and right-hand boundaries corresponds to zero residual
emission as this region has been undisturbed by the jet.
Note the enhanced emission in the region behind the bow-shock and the 
multiple, smaller shocks within the cocoon, as well as the x-ray deficit
in the region where the jet has disrupted. The dark lines along the
left and right-hand sides indicate the positions of the slices shown
in Fig.~\ref{xrayslices}.}
\label{xray_af}
\end{figure}

\begin{figure}[p]
\plotone{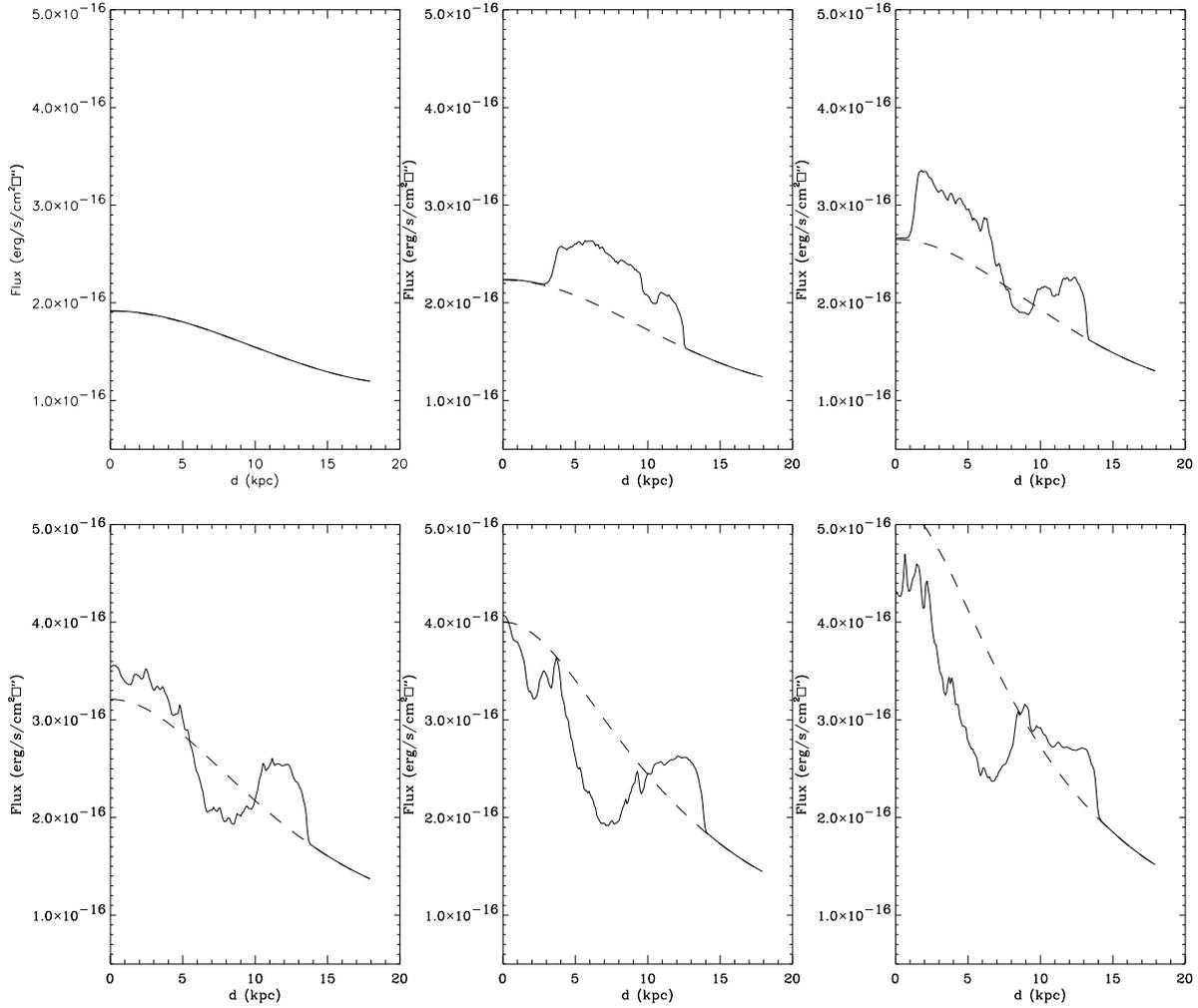}
\caption{Slices of x-ray surface brightness made horizontally
across the synthetic x-ray image at the positions
indicated shown in Fig.~\ref{xray_af}. 
The 6 slices are made at the positions indicated in Fig.~\ref{xray_af}
and are ordered from the top down.
In each case, the solid line is the profile from the final state while
the dashed line represents the corresponding profile from the initial,
steady-state atmosphere.}
\label{xrayslices}
\end{figure}

\begin{figure}[p]
\centerline{
\psfig{figure=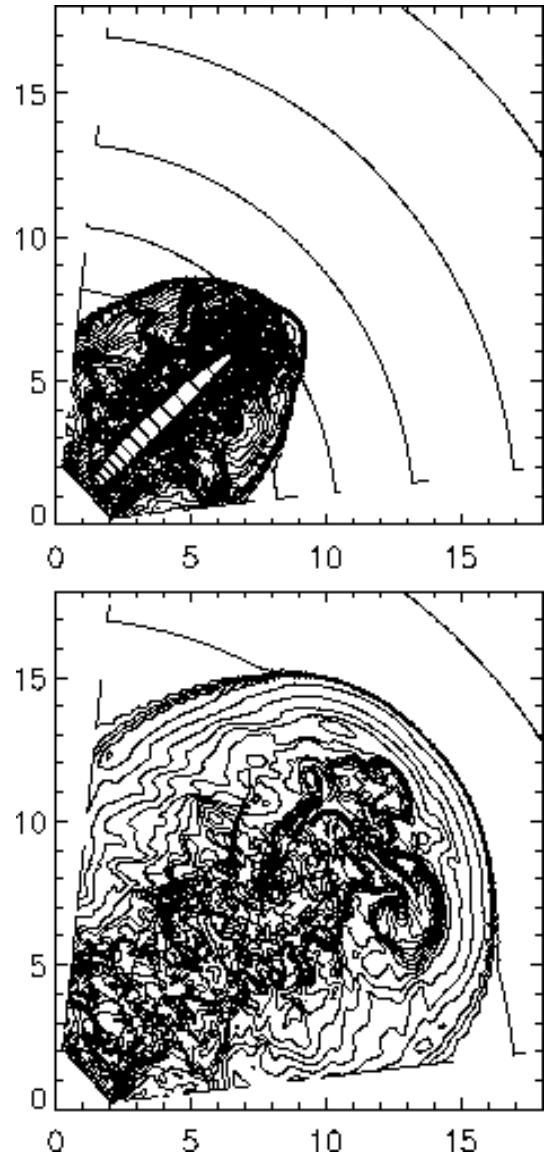,height=6.0in}}
\caption{Slices of log(density) through the mid-plane of the
M$_j$=20 jet just before the inflow at the inlet is turned-off
(top panel; t=1.3$\times 10^{6}$ yrs), and at the final
epoch (bottom panel; t=3.5 $\times 10^{6}$ yrs). }
\label{dens_ak}
\end{figure}

\begin{figure}[p]
\plotone{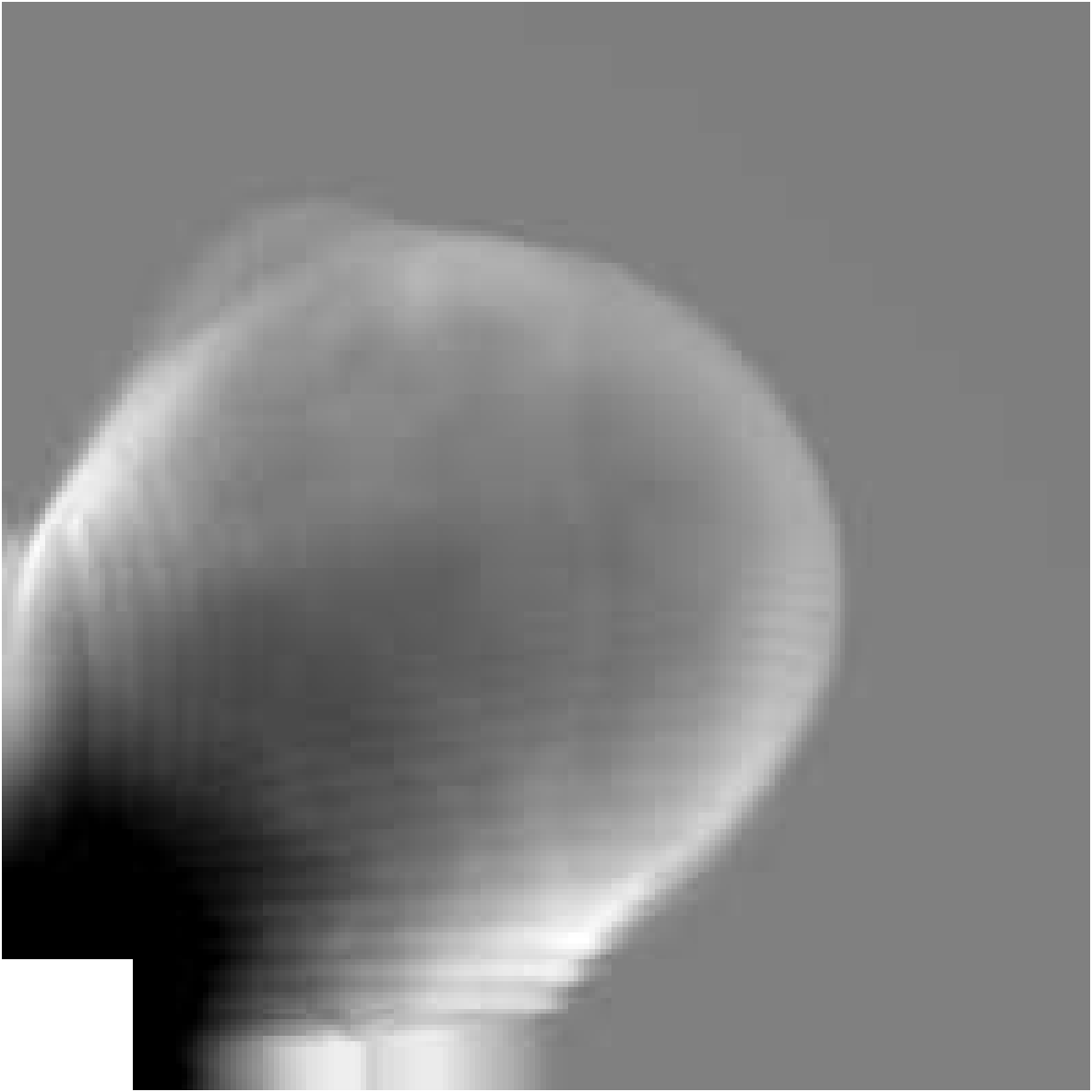}
\caption{Synthetic x-ray image of the M$_j$=20 jet at the final epoch. 
The initial cooling flow atmosphere has been subtracted in order to mimic 
the procedure used to create the observed residual maps. 
Note the enhanced
emission in the region behind the bow-shock as well as the x-ray deficit
in the region where the jet has disrupted.}
\label{xray_ak}
\end{figure}

\end{document}